\documentclass[draftcls,10pt,onecolumn,oneside]{IEEEtran}
\usepackage{mathrsfs}
\usepackage{algorithm}
\usepackage{algorithmic}
\usepackage{epstopdf}
\usepackage{multirow}
\usepackage{booktabs}
\usepackage{threeparttable}
\usepackage{amsmath,amssymb,amsfonts}
\usepackage{graphicx}
\usepackage{textcomp}
\usepackage{enumerate}
\usepackage{subfigure}
\usepackage{mathrsfs}
\usepackage{amssymb}
\usepackage{url,times,amsmath,color,amssymb,graphicx,epsfig,psfig,cite,geometry,psfrag,subfigure,dsfont,booktabs, multirow}
\begin{document}
\title{Electromagnetic Wave Property Inspired Radio Environment Knowledge Construction and AI-based Verification for 6G Digital Twin Channel}

\author{
Jialin Wang, Jianhua Zhang, Yutong Sun, Yuxiang Zhang, Tao Jiang, Liang Xia

\thanks{
Jialin Wang, Jianhua Zhang, Yutong Sun, Yuxiang Zhang are with State Key Laboratory of Networking and Switching Technology, Beijing University of Posts and Telecommunications, Beijing 100876, China (e-mail: wangjialinbupt@bupt.edu.cn, jhzhangn@bupt.edu.cn, sun\_yutong@bupt.edu.cn, zhangyx@bupt.edu.cn)
Tao Jiang and Liang Xia are with China Mobile Research Institution, China (e-mail:jiangtao@chinamobile.com, xialiang@chinamobile.com).

$*$  \quad Correspondence: jhzhangn@bupt.edu.cn
}
}

\maketitle

\begin{abstract}
As the underlying foundation of a digital twin network (DTN), a digital twin channel (DTC) can accurately depict the process of radio propagation in the air interface to support the DTN-based 6G wireless network.
Since radio propagation is affected by the environment, constructing the relationship between the environment and radio wave propagation is the key to improving the accuracy of DTC, and the construction method based on artificial intelligence (AI) is the most concentrated.
However, in the existing methods, the environment information input into the neural network (NN) has many dimensions, and the correlation between the environment and the channel relationship is unclear, resulting in a highly complex relationship construction process.
To solve this issue, in this paper, we propose a construction method of radio environment knowledge (REK) inspired by the electromagnetic wave property to quantify the contribution of radio propagation.
Specifically, a range selection scheme for effective environment information based on random geometry is proposed to reduce the redundancy of environment information. 
We quantify the contribution of radio propagation reflection, diffraction and scatterer blockage using environment information and propose a flow chart of REK construction to replace the feature extraction process partially based on NN.
To validate REK's effectiveness, we conduct a path loss prediction task based on a lightweight convolutional neural network (CNN) employing a simple two-layer convolutional structure.
The results show that the accuracy of the range selection method reaches 90\%; the constructed REK maintains the prediction error of 0.3 and only needs 0.04 seconds of testing time, effectively reducing the network complexity.

\end{abstract}

\section{Introduction}

With the ever-increasing demand for wireless communication, the sixth generation (6G) wireless network will be more complex \cite{zhang2022toward,rajoria2022brief,giordani2020toward}.
Meanwhile, it needs to meet the needs of large-scale users, expand new services, applications and scenarios, and add new techniques such as cloud-native and information technology virtualization  \cite{chowdhury20206g,Liu2023digital,saad2019vision,liu2020vision,ji2022discussion,zhang2020ubiquitous}. 
The digital twin network (DTN) provides up-to-date network status, closed-loop decision-making, and real-time interaction between the digital and physical worlds by creating a real-time digital replica of physical entities, which adapts the complex 6G wireless network \cite{khan2022digital,lin20236g,wu2021digital,khan2022digital02,masaracchia2022digital}.
To realize the DTN, it is essential to acquire the accurate wireless channel states between transmitters (TXs) and receivers (RXs), including propagation parameters, fading conditions and other information related to channels. 
Therefore, the concept of digital twin channel (DTC) is proposed that enables the mapping of physical world channel states into the digital worlds, emphasizing realized interaction between channel and system under the environment changes \cite{wang2023towards,wang2024digital}.  
As the underlying foundation of DTN, DTC enhances the accuracy of real-time channel representation and supports more reliable communication transmission and network optimization for DTN-based 6G wireless networks.

\begin{figure*}[tbh]
	\centering
	\includegraphics[scale=0.38]{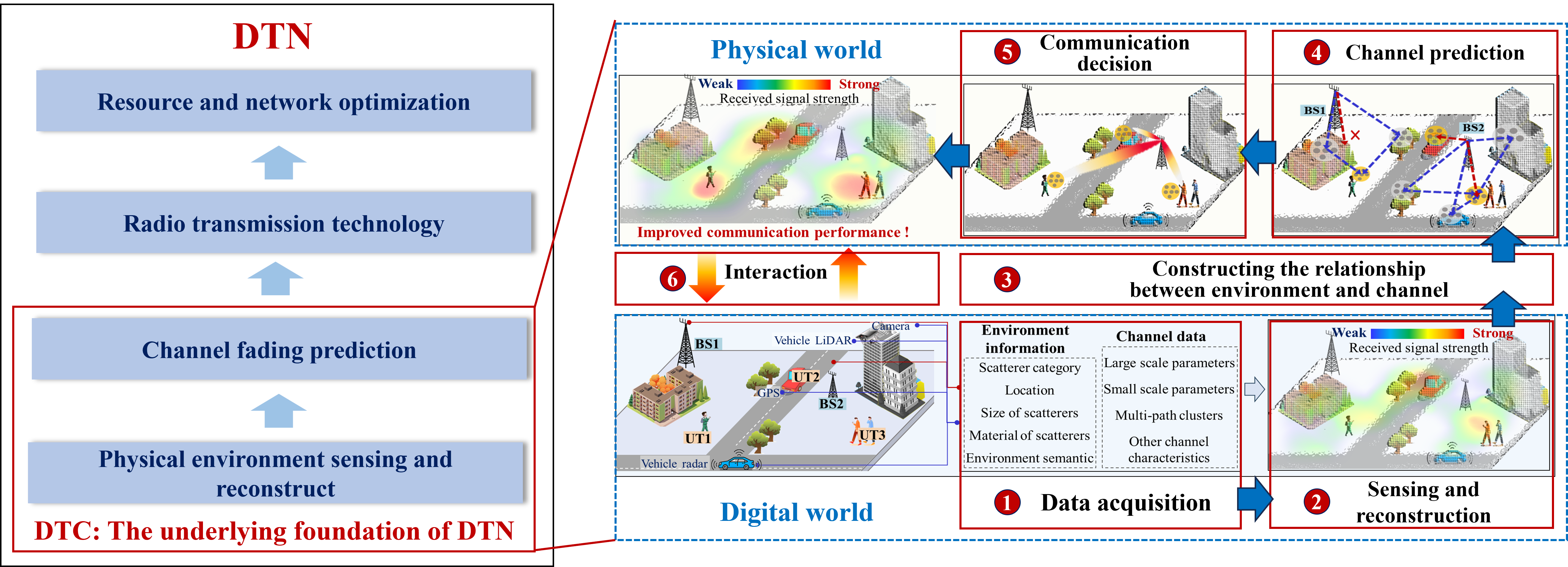}
	\caption{The implementation framework of DTC includes six parts.} \label{DTC}
\end{figure*}

The 6G overall objectives and framework proposed by the International Telecommunication Union (ITU) identify six major scenarios, with integrated sensing and communication and integrated artificial intelligence (AI) and communication being two of them \cite{re2023framework,wp5d2022future}.
The foreseeable development of sensing and AI techniques strongly supports the implementation of DTC \cite{zhang2021intelligent,wang2022intelligent,liu2022integrated}.
The implementation framework is shown in Fig. \ref{DTC}. 
Firstly, environment information and channel data are captured from the physical world for environment sensing and reconstruction, which involves channel measurement \cite{zhang2020channel,rappaport2019wireless,zhang2023channel}, environment-channel datasets construction \cite{shen2023dataai,cheng2023m}, environment reconstruction \cite{8618413,8736013}.
Then, channel prediction and communication decisions are performed in the digital world by constructing the relationship between environment and channel, which involves channel large/small-scale parameter prediction \cite{hayashi2023deep,sun2022environment}, low overhead channel state information (CSI) prediction \cite{hou2021federated,zhang2023deep}, decision of beam selection and power allocation \cite{wu2023environment,nguyen2023power}.  
Finally, the decision result is fed back from the digital world to the physical world, and the real-time propagation environment information in the physical world is combined to realize the interaction.

Constructing the relationship between environment and channel is key to the six parts because this part is the bridge to achieving accurate channel presentation in a changing environment.
Some studies focus on introducing environment information into channel representation based on traditional theory, such as the cluster-nuclei-based channel modeling \cite{zhang2016interdisciplinary}, the channel twin platform based on global environment sensing \cite{miao2023demo}, the channel parameter optimization based on building height \cite{bor2017environment}, path loss modeling based on point cloud environment data \cite{jarvelainen2016evaluation}.
Recently, some studies on obtaining feature matrices representing the relationship between environment and channel directly based on AI.
For example, a graph neural network is used to obtain the relationship between the relative location of TX, RX and scatterers in relation to channel large/small scale parameters \cite{sun2022GNN}.
The relationship between the environment images of different terminal locations and the corresponding channels is constructed through deep learning \cite{yang2019deep}
The relationship between environment semantics and beam \cite{yang2023environment,zhang2023model}, environment images and CSI \cite{Wang2020Deep} are studied based on deep neural networks.
High-dimensional environment information is required to obtain the relationship between the environment and the channel in complex scenarios. 
To meet performance requirements, complex neural network structures for deep feature extraction are needed. 
However, no research supports which environment information in one scene affects the electromagnetic wave propagation process, and how to quantify the influence trends and degree. 
This is one of the ways to characterize the essential relationship between the environment and the channel, which can be realized by constructing a radio environment knowledge pool (REKP) that we mentioned recently \cite{wang2024digital}.
The essential relationship provides the basis for relationship construction and removes environment information redundancy, which is conducive to simplifying the neural network structure and establishing relationships quickly to support DTC.

As we all know, electromagnetic wave propagation is influenced by the environment. 
Electromagnetic waves interact with buildings, trees, vehicles, and other environments through reflection, diffraction, and direct radiation, the propagation path of which forms a geometric relationship between TX and RX in the scene.
Inspired by this, this paper fills the gap in the quantitative analysis of the relationship between effective information in the environment and electromagnetic wave propagation. 
We propose a construction method of radio environment knowledge (REK) that uses environment information to quantify the contribution of direct, reflection and diffraction. The main contributions are as follows:

\begin{itemize}
    \item A range selection scheme for effective environment information based on random geometry is proposed, and the range selection accuracy reaches 90\% in the case of impending blockage and complete blockage.
    
    \item A reflection point acquisition algorithm based on random scatterer shortest path is proposed, and the contribution of reflection, diffraction and blockage are quantified using environment information.

    \item The impact of the geometric topology of the environment on electromagnetic wave propagation is analyzed, and the implementation process of the REK is proposed, including the situation of complete openness, impending blockage, and complete blockage.
    
    \item To verify the constructed REK, a lightweight CNN-based path loss prediction method is proposed using a simple CNN structure with two convolutional layers. The results show that the path loss prediction error is 0.3 with only 5 and 0.004 seconds of training and testing time.
\end{itemize}

\section{Problem formulation}

\begin{figure*}[tbh]
	\centering
	\includegraphics[scale=0.5]{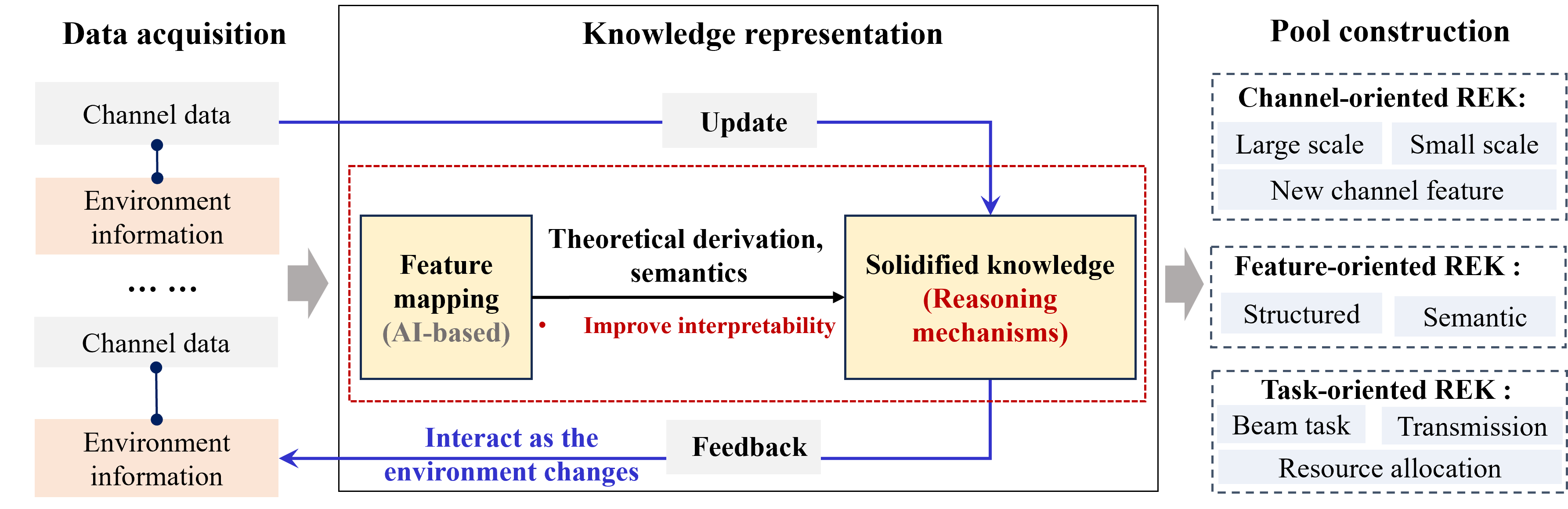}
	\caption{REKP architecture with its core function being the construction of radio environment knowledge.} \label{fig:rekp}
\end{figure*}

\subsection{Scene and assumption}
We consider a sensing-assisted communication single-cell scene. 
One single base station (BS) serves $R$ user terminals (UTs), employing some sensing devices.
The BS acquires the following information: 
(1) Environment data of stationary objects (such as buildings, trees and so on.) are provided in the form of images or values, including their locations, materials, sizes, and so on.
(2) Location information of fast-moving objects acting as reflective entities that do not communicate with the BS.
(3) Location and channel information of UTs communicating with the BS. Various propagation states exist between the base station and terminals, including line-of-sight (LoS), reflection, scattering, and blockage.

In communication systems, the object that causes the signal to scatter, attenuate or reflect is called a scatterer.
Scatterers have random shapes, especially buildings in outdoor urban areas and indoor office areas, which are densely distributed and display various shapes.
Given the correlation between signal propagation and scatterers and the interrelationship between channel quality and scatterer information such as shape and location, subject to the following assumptions:

The scatterers are modeled as the line Boolean model \cite{bai2014analysis}. 
Scatterers are abstracted into varying-sized cuboids, represented by the length of the diagonal segments. 
The central locations of all line segments are modeled by ${\Phi _s} = \left\{ {{P_i}} \right\} \in \mathbb{R}^2$, where ${{P_i}}$ represents the location of the center point of the $i$-th scatterer, and the length of each line segment is denoted as ${L_i} = d({P_{\min }},{P_{\max }})$, follows a uniform distribution ${L_i} \sim {\bf{U}}\left[ {{L_{\min }},{L_{\max }}} \right]$ with a mean value of ${\mathbb{E}}[L]$.
$d\left(  \cdot  \right)$ represents the calculation of the Euclidean distance, ${{P_{\max }}}$ and ${{P_{\min }}}$ represent the minimum and maximum coordinates within a scatterer, and they are obtained through sensing devices.
UTs are considered stationary within short time intervals, with their locations characterized by ${\Phi _u} = \left\{ {{P_r}} \right\} \in \mathbb{R}^2$, where ${{P_r}}$, and ${\Phi _u}$ represent the location coordinates of the $r$-th user, and the set comprising all user location coordinates, respectively.

It is assumed that the data transmission links between base stations and users currently employ existing beamforming techniques, effectively mitigating interference among multiple users within the same cell.   
In other words, interference terms can be disregarded in a single-cell millimeter-wave wireless communication system.


\subsection{Radio environment knowledge pool (REKP)}
To implement DTC, we recently proposed the REKP to characterize and store diverse environment and channel relationships \cite{wang2023towards}, as shown in Fig. \ref{fig:rekp}.
Its core functionality lies in using environment information to explore the interpretable relationships (the essence, interconnections, and principles) between channel information and the environment information and form solidified REK.
Interesting channel information includes but is not limited to channel large or small-scale parameters, CSI, channel propagation, communication tasks affected by channel characteristics, and other information related to channel parameters.
The environment information emphasizes using multi-modal, multi-type and large amounts of environment data to describe one communication scene comprehensively.
Unlike previous studies emphasizing feature mapping obtained through AI algorithms, REKP focuses on generating solidified knowledge through reasoning mechanisms such as theoretical derivation or semantic construction, thereby improving the interpretability of the channel-environment relationship.

REKP viewes as a collection of interpretable relationship mappings $K\left(  \cdot  \right)$ from environment information $\Phi  \in {\mathbb{R}^D}$ to channel knowledge vectors $k  \in {\mathbb{C}^J}$, expressed as:
\begin{equation}
\mathcal{K}= \left\{ {K\left(  \cdot  \right):{\mathbb{R}^D} \to {\mathbb{C}^J}} \right\}
\end{equation}
where $\Phi  \in {\mathbb{R}^D}$ represents a collection of $D$-dimensional environment information obtained in a physical world.
$\mathbb{R}$ includes environment data about BS, UTs and scatterers, such as their locations, volumes, materials, antenna height, and so on.
$D$ is scenario-related, varying with different types of scenarios.
For example, blockage obstructed in the non-line-of-sight (NLoS) condition is considered, resulting in a larger dimensionality of environment information in NLoS situations compared to the LoS condition.
$k  \in {\mathbb{C}^J}$ denotes the REK that can adapt to environment changes, where $J$ depends on the channel requirements of communication systems, such as channel-oriented REK, feature-oriented REK, communication task-oriented REK, and so on.
$\mathcal{K}\left(  \cdot  \right)$ represents the process of REK expression, which constructs the interpretable relationships between the environment and the channel.
After quantifying the interpretable relationship through a large amount of environment information, it exhibits scene generalization without considering the specific characteristics of the communication scene.
 
\section{REK construction}

\begin{figure}[tbh]
	\centering
	\includegraphics[scale=0.4]{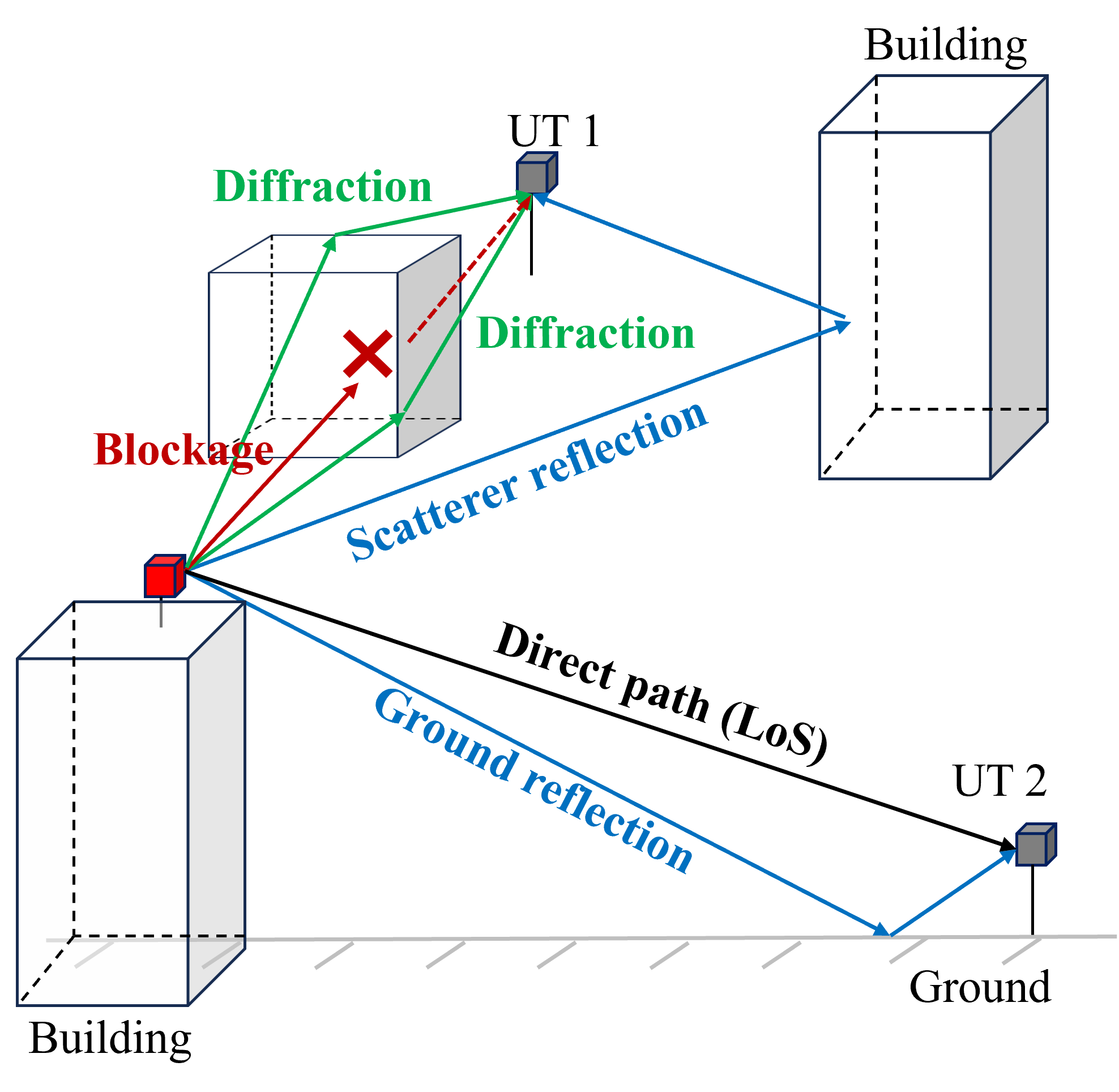}
	\caption{The radio propagation process from BS to UT includes three types: direct, reflection, diffraction.} \label{3Propagation}
\end{figure}

Due to the limitation of antenna height and the influence of surroundings, build-up urban areas rely more heavily on the environment around the antennas.
Antennas at the BS are usually installed at rooftop heights. 
Affected by diffracted signals around building corners, antenna height restrictions, and blockages caused by building layout, the propagation process from BS to UTs can be viewed as the signal originating from the TX, passing through building tops, facades, or edges, nearby buildings, and reaching UTs. 
This process involves three modes of radio propagation, as illustrated in the Fig. \ref{3Propagation}:
(1) Direct radio propagation to the UTs, where the user is within the LoS range for free space propagation. 
(2) Reflection radio propagation from building walls and ground to UTs.
(3) Diffraction radio propagation comes from the sides of the building and the roof of the building near UTs.

The wireless channel is generated through various physical propagation types such as direct radiation, reflection, and diffraction of electromagnetic waves. The channel model can be represented as:
\begin{equation}
h\left( t \right) = {h_{los}}\left( t \right) + {h_{ref}}\left( t \right) + {h_{df}}\left( t \right)\
\end{equation}
\begin{equation}
{h_i}\left( t \right) = \sum\limits_{n = 1}^{{N_i}} {{\beta _{n,i}}\left( t \right)} {e^{j\left( {2\pi {v_{i,n}}t + {\phi _{i,n}}\left( t \right)} \right)}}
\end{equation}
where $i$ is the propagation type, that is, direct radiation, reflection, and diffraction. 
${N_i}$ is the number of channel multipaths generated by types $i$. ${\beta _{n,i}}\left( t \right)$, ${\phi _{i,n}}\left( t \right)$, and ${v_{i,n}}$ represent time-varying path loss, phase, and Doppler shift, respectively.
Next, the REK is constructed from environment information, delving into the essence of the relationship between environment and radio propagation.

\subsection{Overall construct flow}

\begin{figure*}[tbh]
	\centering
	\includegraphics[scale=0.35]{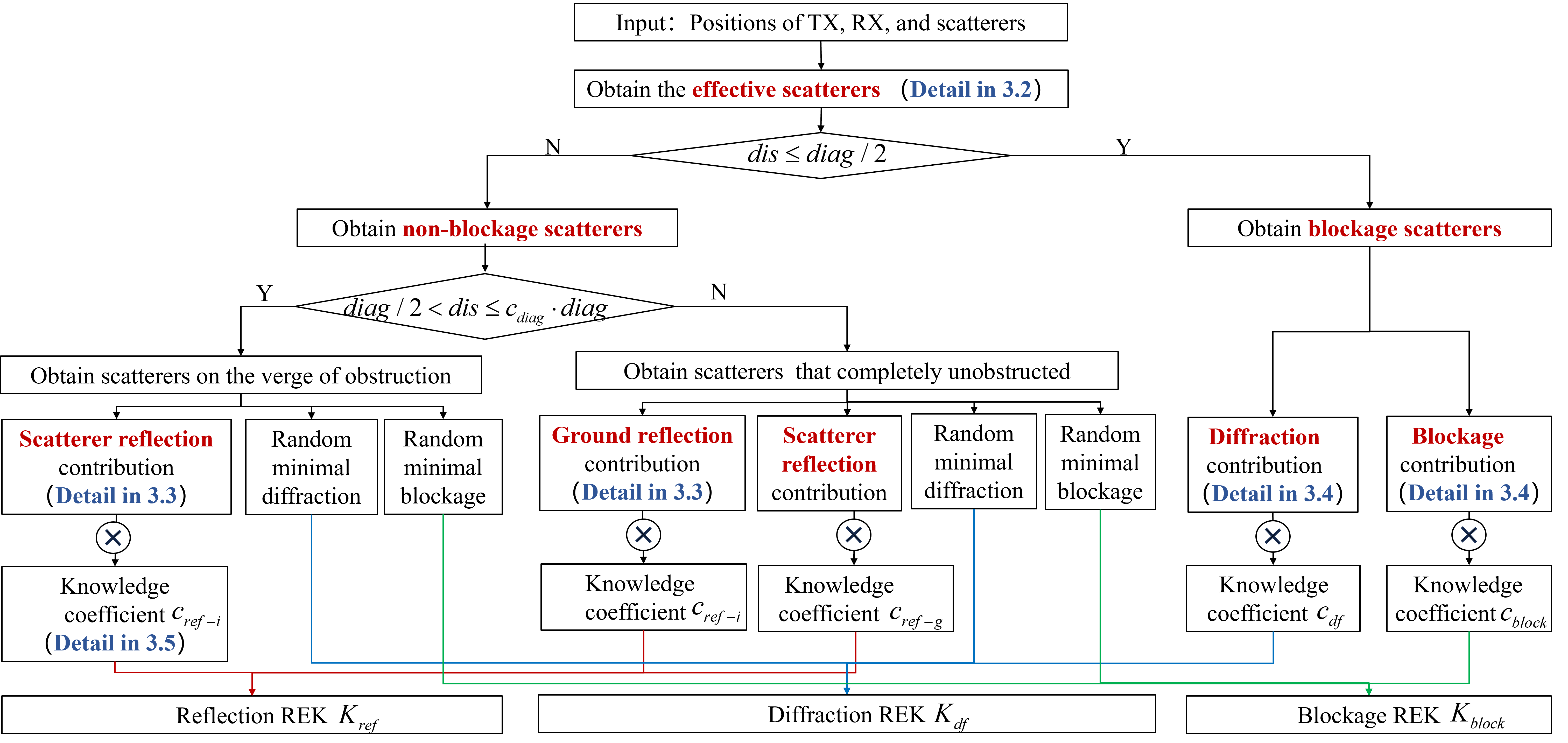}
	\caption{The flow diagram of the REK construction for one UT.} \label{PREKflow}
\end{figure*}

The values of path gain are determined by the superposition of multiple subpaths, which are generated based on the environment structure between the TX and RX. 
Therefore, the constructed REK ${k_{ppg}}$ consists of two components: a representation of propagation knowledge based on environment information ${\bf{M}}$, and knowledge coefficients ${\bf{C}}$, expressed as:
\begin{equation}
{\bf{K}}\left( {{\bf{M,C}}} \right):\left\{ {{P_i},{P_r},{P_t}} \right\} \to {k_{ppg}}
\end{equation}
where ${{P_i}}$, ${{P_r}}$, and ${{P_t}}$ are the positions of the $i$-th scatterer, the $r$-th RX, and the TX, respectively.
The propagation knowledge explains the contribution levels of propagation modes in the current scene based on environment information, and the knowledge coefficients reflect the weights of propagation modes in different scenes.
The flow diagram for the REK construction is shown in Fig. \ref{PREKflow}.

\textbf{\emph{Step 1:}}
Using environment information, the ellipsoid model is constructed based on the stochastic geometry and the range of effective scatterers involved in radio propagation is determined.

\textbf{\emph{Step 2:}}
Under the current user location, determine the blockage scatterers that affect LoS propagation and save the corresponding number. 
If there is a scatterer $S$, where the distance from its center $O$ to $d_{t, r}$ is less than half of the scatterer's diagonal length, then this scatterer $S$ is classified as a blockage scatterer, denoted as:
\begin{equation}
B = \left\{ {\exists {P_i},\frac{{{{\bf{P}}_i} \cdot \overrightarrow {{{\bf{P}}_t}{{\bf{P}}_r}} }}{{||\overrightarrow {{{\bf{P}}_t}{{\bf{P}}_r}} ||}} < \frac{{{L_i}}}{2}|i \in S} \right\}
\end{equation}
where ${{L_i}}$ is the size of the $i$-th scatterer.
Those that are not judged as blockage scatterers are classified as non-blockage scatterers.

\textbf{\emph{Step 3:}}
Under the current user location, determine the scatterers that will become blockage scatterers in the range of non-blockage scatterers and save the corresponding numbers.
The others are divided into completely openness situations that do not affect LoS.

\textbf{\emph{Step 4:}}
For scatterers on the verge of obstruction, their reflective contribution is computed based on the light reflection theorem in geometric optics (GO) theory \cite{katz2002introduction}, and acquiring a knowledge coefficient ${\bf{C}}$.
In the situation of approaching occlusion, diffraction and blockage contributions are minimal. 
Here, their contributions are assigned random values between 0 and 1. 
Repeat \textbf{\emph{Step 4}} until all the propagation knowledge that will become the blockage scatterers is obtained.

\textbf{\emph{Step 5:}}
In the LoS propagation mode, the superlocation of ground reflection and direct waves at the receiving antenna leads to multipath effects. 
Hence, the contribution of ground reflection cannot be overlooked in LoS scenarios. 
For scatterers entirely exempt from becoming occlusions, the ground reflection contribution is quantified based on the dual-line reflection model of GO.

\textbf{\emph{Step 6:}}
For blockage scatterers, diffraction propagation losses are quantified based on the uniform geometrical theory of diffraction (UTD) \cite{pathak2011uniform} and the Fresnel principles \cite{saunders2007antennas}. 
The blockage contribution in the propagation process is quantified based on geometric relationships.

\textbf{\emph{Step 7:}}
Finally, the contributions of reflection, diffraction, and blockage obtained in the three situations are consolidated to obtain the REK towards radio from BS to one UT.

\subsection{Effective and blockage scatterers}\label{3.2}

\begin{figure*}[tbh]
	\centering
	\includegraphics[scale=0.5]{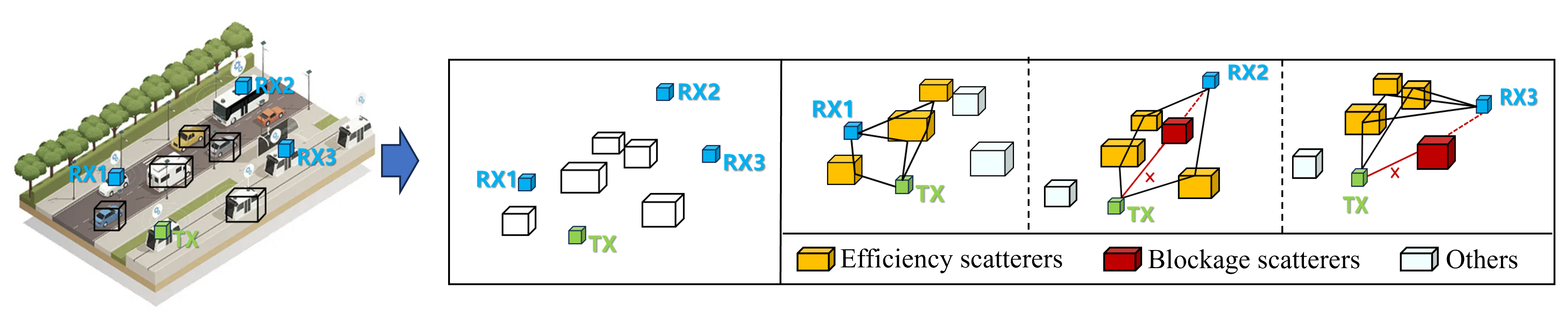}
	\caption{Effective, blockage, and other scatterers in the communication scenario.} \label{scatterers}
\end{figure*}

\begin{figure}[tbh]
	\centering
	\includegraphics[scale=0.33]{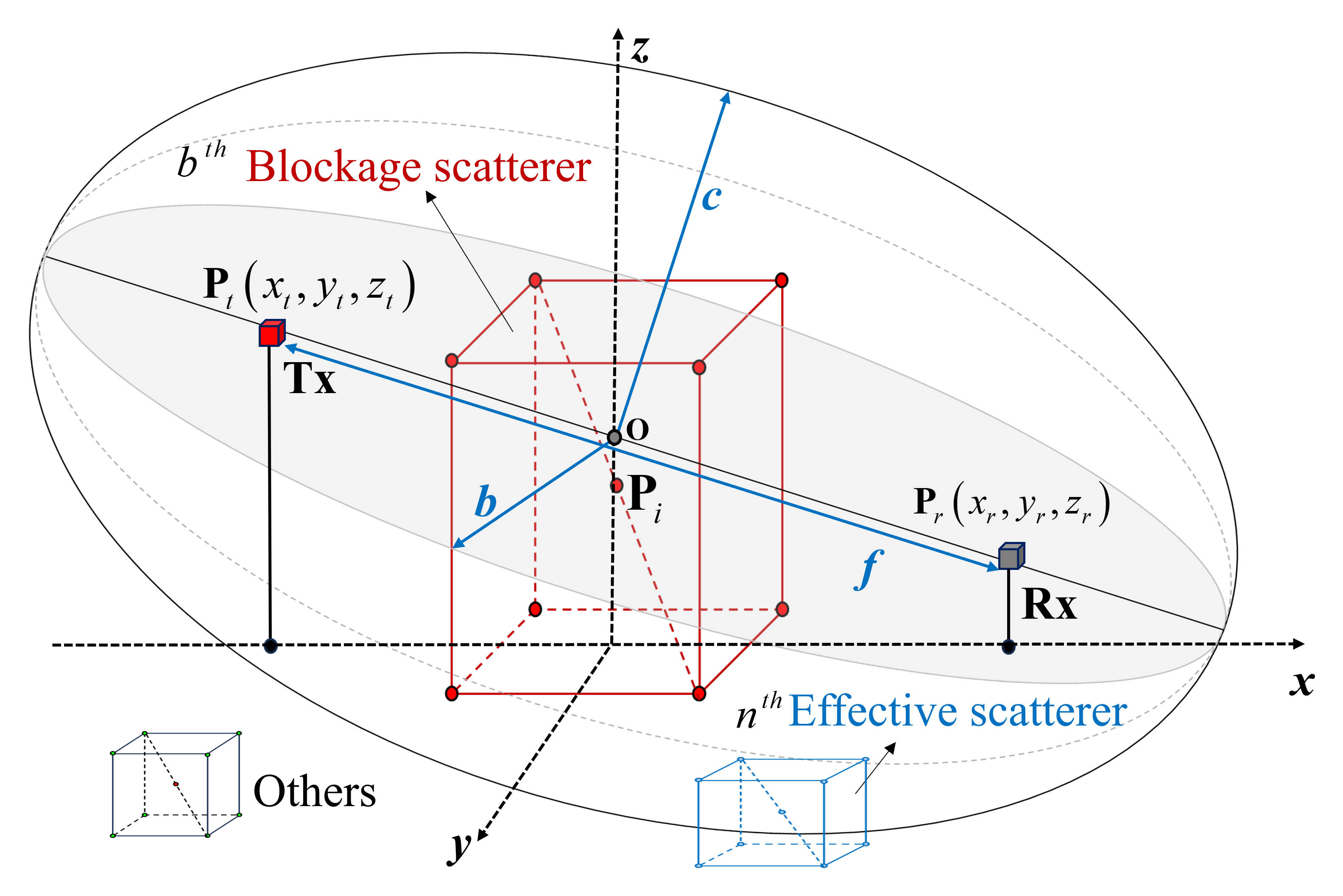}
	\caption{The scatterer ellipsoidal model based on stochastic geometry.} \label{ellipsoidal}
\end{figure}

We divide scatterers in the communication scenario into effective, obstructive, and other scatterers, as shown in Fig. \ref{scatterers}. 
Effective scatterers refer to the scatterers that the wireless signal passes through in the transmission process from BS to UT in the propagation mode of reflection, scattering, and so on.
Blockagescatterers block the signal from TX so that Rx cannot accept the signal.
In the REK construction process, the same building is judged as a different class of scatterers in the face of different UTs or as the UT moves, related to the user location and the geometric path among TX, RX and scatterers.

The scatterer ellipsoidal model based on stochastic geometry is established as shown in Fig. \ref{ellipsoidal}. 
TX and RX locations are focal points, and the half-axis lengths along the three orthogonal axes (usually x, y, and z-axis) are denoted as $a$,$b$, and $c$, respectively. The focal distance $f$ is the distance $d_{t, r}$ between TX and RX, $O\left( {{x_0},{y_0},{z_0}} \right)$ is the center point coordinate of the ellipsoid.
It is assumed that there are $S$ scatterers in the scene, including $N$ effective, $B$ obstructive, and other scatterers. 

Assuming $b = c = \frac{1}{2}f$ and $a = \sqrt {2{b^2}} $, there exists a scatterer $S$ with vertex coordinates from ${P_{b1}}$ to ${P_{b8}}$.
If all eight vertex coordinates are located within the ellipsoid, or if one of the six surfaces of scatterer $S$ intersects with the ellipsoid, then this scatterer is classified as an efficient scatterer, denoted as:

\begin{footnotesize}
   \begin{equation}
N = \left\{ {\exists {P_i}\left( {{x_i},{y_i},{z_i}} \right),\frac{{2{x_i}^2}}{{{d_{t,r}}^2}} + \frac{{4{y_i}^2}}{{{d_{t,r}}^2}} + \frac{{4{z_i}^2}}{{{d_{t,r}}^2}} < 0|i \in S} \right\}
    \end{equation} 
\end{footnotesize}

It is worth mentioning that the calculated effective scatterers include scatterers deploying TX and RX. However, in practice, this situation is excluded from the set of effective scatterers.

\subsection{Reflection contribution}\label{3.3}
When radio waves pass from one medium to another, such as from one medium to another medium, a portion reflects, obeying the law that the angle of reflection equals the angle of incidence.
We first propose a reflection point acquisition algorithm based on a random scatterer's shortest path. This algorithm can retrieve reflection points solely through the locations of transmitters, receivers, and scatterers in a random scenario, as illustrated in Algorithm \ref{alg:update-alg}.
\begin{algorithm}\small
\centering
\caption{Reflection point acquisition algorithm based on random scatterer shortest path} \label{alg:update-alg}
\begin{algorithmic}[1]
	\STATE {$Faces,Normals \leftarrow {P_i}$}
	\LOOP	
            \STATE {${D_{\min }} = \infty $}
		\FOR{ $i \gets  1 ~\mathrm{to}~ num(Faces) $}
		\STATE{$TS \leftarrow Faces\left[ i \right],{P_t}$}
  		\STATE{$SR \leftarrow Faces\left[ i \right],{P_r}$}
  		\STATE{${D_{sum}} = TS + SR$}
            \IF{${D_{sum}} < {D_{\min }}$}
                \STATE{update to ${D_{min}}$:${D_{\min }} = \min \left( {{D_{sum}}} \right)$}
                \STATE{$\mathcal{N} \leftarrow Normals\left[ i \right]$}
                \STATE{$Q \leftarrow Faces\left[ i \right]\left[ 0 \right]$}
		\ENDIF
            \STATE{$\hat P \leftarrow {P_t},\mathcal{N},Q$}
            \STATE{$RP \leftarrow {P_r},\mathcal{N},Q,\hat P$}
	\ENDFOR
        \ENDLOOP
\end{algorithmic}
\end{algorithm}

\begin{figure}[tbh]
	\centering
	\includegraphics[scale=0.4]{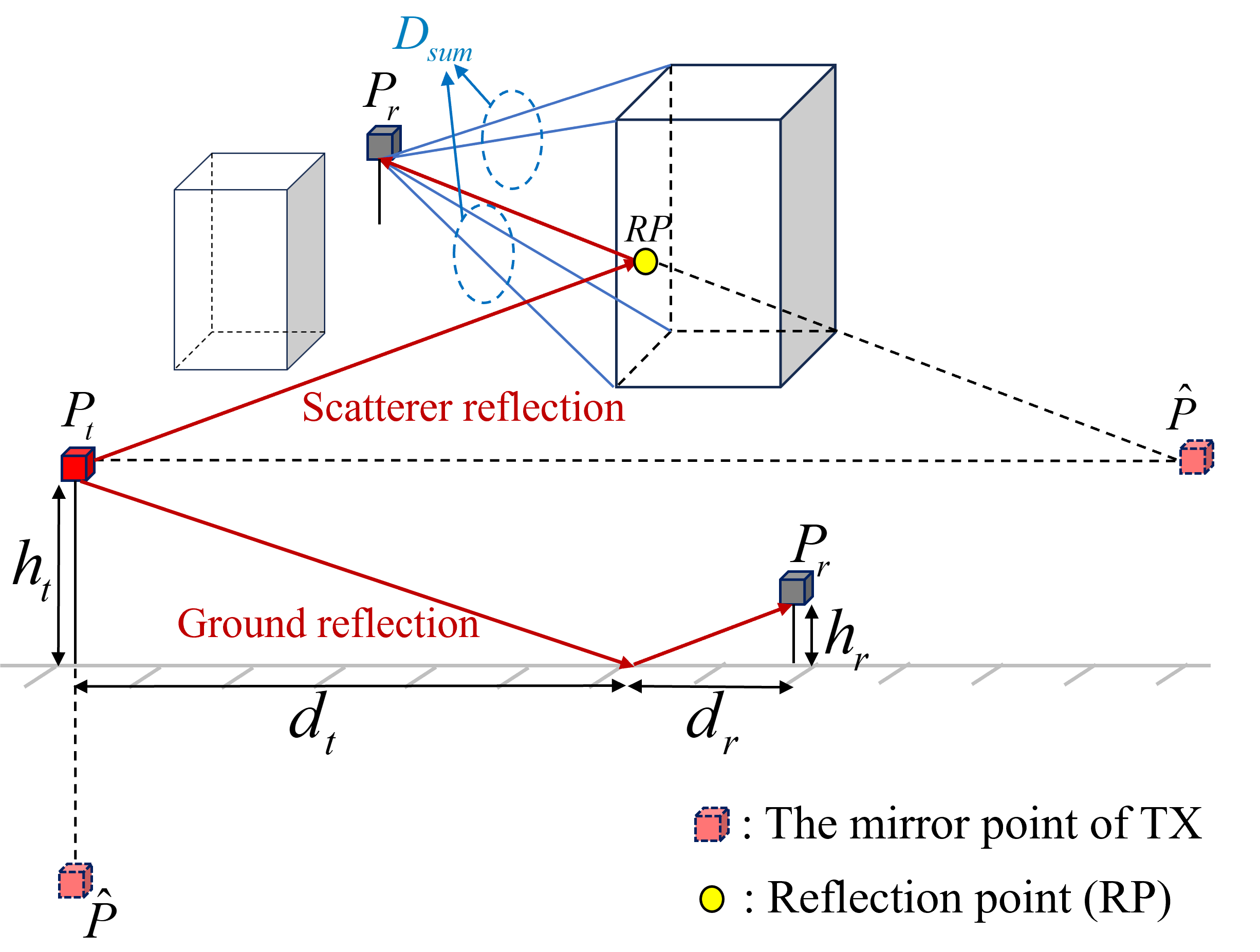}
	\caption{The geometric relationship for reflection contribution.} \label{reflection}
\end{figure}

A face is calculated every four vertices through the center point $P_i$, totaling six faces. 
The coordinates of the four points of each face are stored as a group, totaling six groups stored in $Faces = \left\{ {{F_1},{F_2}, \cdots, {{F_\varsigma }}} \right\},\varsigma  = 1,..,6$. 
Based on two lines perpendicular to each other on each face, the normal vector of each face is obtained, totaling six normal vectors stored in $Normals = \left\{ {No{r_1},No{r_2}, \cdots ,No{r_\varsigma }} \right\},\varsigma  = 1,..,6$. 
The sum of the distances from the four vertices on each face to TX and RX is expressed as:
\begin{equation}
{D_{sum}} = \sum\limits_{p = 1}^4 {\left\| {{v_p} - {P_t}} \right\| + } \left\| {{P_r} - {v_p}} \right\|\left( {{v_p} \in Faces} \right)
\end{equation}
where ${{v_p}}$ represents the $p$-th vertex on a face, $\left\|  \cdot  \right\|$ is the norm operation.
The face where the smallest ${D_{sum}}$ is located is the face where the reflection point is located. The specular reflection point of TX to this surface is given:
\begin{equation}
\hat P = {P_t} - \left\langle {{P_t} - Q,\frac{\mathcal{N}}{{\left\| \mathcal{N} \right\|}}} \right\rangle  \cdot \frac{{2\mathcal{N}}}{{\left\| \mathcal{N} \right\|}}
\end{equation}
where $\mathcal{N}$ is the normal vector of the plane, $Q$ is any point on the plane, and $\left\langle  \cdot  \right\rangle $ is vector dot product operation. 
In the proposed algorithm, $Q$ is taken as the first vertex in the plane.
Then, the reflection point is expressed as:
\begin{equation}
RP = \hat P + \frac{{\left\langle {Q - \hat P,\hat N} \right\rangle  \cdot \left( {{P_r} - \hat P} \right)}}{{\left\langle {{P_r} - \hat P,\hat N} \right\rangle }}
\end{equation}
where $\hat N$ is normalizing the normal vector $\mathcal{N}$ to a unit vector.
Therefore, the reflection contribution from the scatterer and the ground are expressed as:
\begin{equation}
{K_{ref - i}} = \frac{{{c_{ref - i}}\left\| {{P_t} - {P_r}} \right\|}}{{\left\| {{P_r} - RP} \right\| + \left\| {RP - {P_t}} \right\|}}
\end{equation}
\begin{equation}
{K_{ref - g}} = \frac{{{c_{ref - g}}\left\| {{P_t} - {P_r}} \right\|}}{{\sqrt {{{\left( {{d_t} + {d_r}} \right)}^2} + {{\left( {{h_t} + {h_r}} \right)}^2}} }}
\end{equation}
where ${{c_{ref - i}}}$ and ${{c_{ref - g}}}$ represent the knowledge coefficients of the scatterer and ground reflection, respectively. ${d_t}$ and ${d_r}$ are the horizontal distances from TX and RX to the ground reflection point, and ${h_t}$ and ${h_r}$ are the antenna heights of TX and RX, respectively.
The geometric relationship between TX, RX, the scatterer, and the ground for reflection contribution is shown in Fig. \ref{reflection}.

\subsection{Diffraction and blockage contribution}\label{3.4}

Diffraction refers to the phenomenon where radio waves deviate from their original straight-line propagation path when encountering obstacles during propagation. However, the main signal is not completely blocked and can still be received by users.
Along the actual radio wave propagation path, when a building begins to fall within the first Fresnel zone, diffraction losses occur, leading to decreased received power and reduced signal quality. 
The distance from the vertex of buildings to the line connecting the BS and the UTs is known as the Fresnel clearance.
The convention is that the Fresnel clearance is positive when obstructed, and negative when unobstructed.

\begin{figure}[tbh]
	\centering
	\includegraphics[scale=0.45]{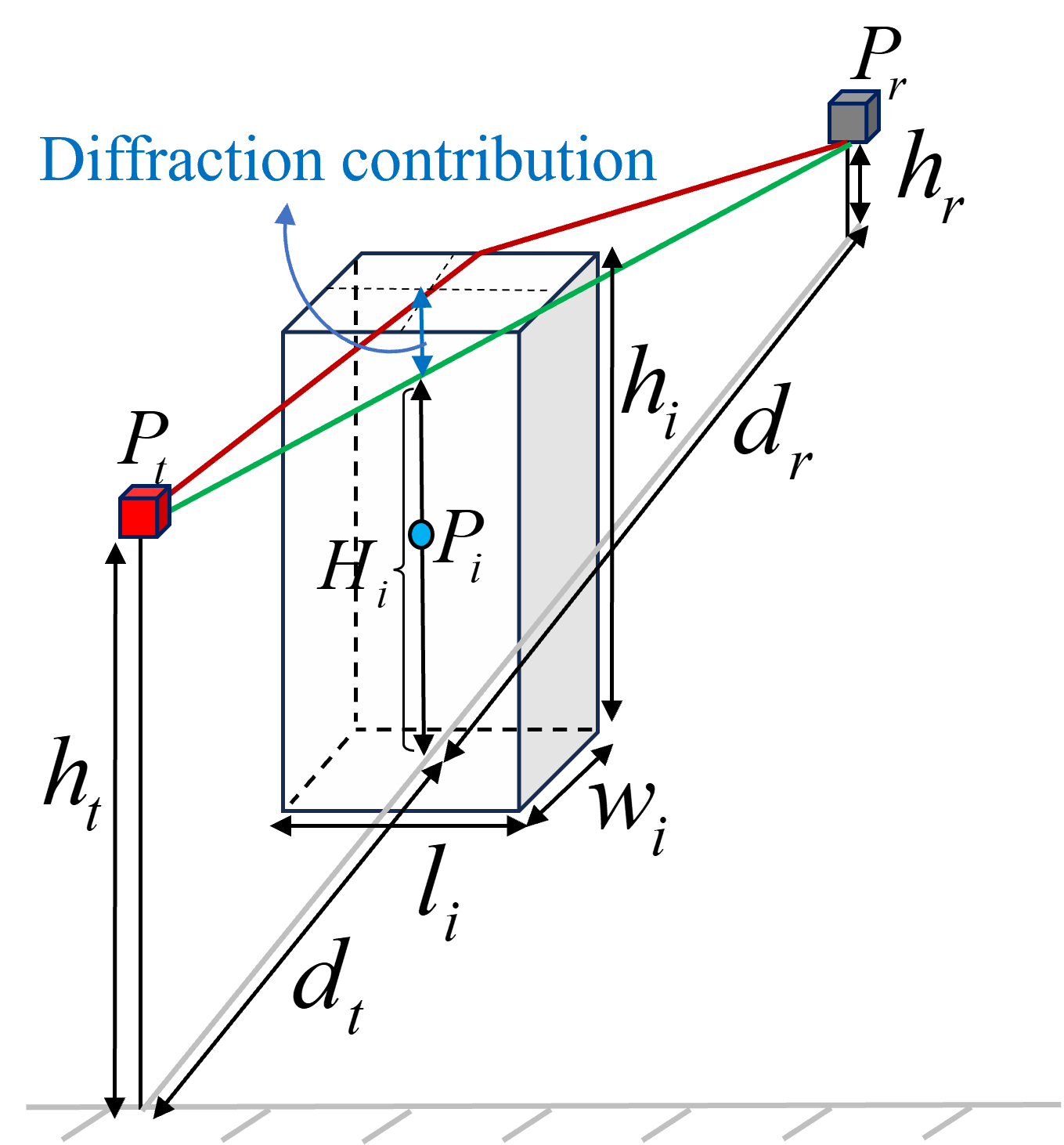}
	\caption{The geometric relationship for diffraction and blockage contribution.} \label{dfandblock}
\end{figure}

Based on this property, a geometric relationship for diffraction and blockage contributions is designed, as shown in Fig. \ref{dfandblock}.
Using the height of the center of the Fresnel zone as a reference, when the height of the obstruction is ${h_i}$, the horizontal distances from the TX and RX to the obstruction are ${d_t}$ and ${d_r}$, respectively, with the transmitting antenna height as ${h_t}$ and the receiving antenna height as ${h_r}$. 
Then the height of the center of the Fresnel zone ${H_i}$ (assuming single scattering) is represented as:
\begin{equation}
{H_i} = {h_r} + \frac{{\left( {{h_t} - {h_r}} \right){d_r}}}{{{d_t} + {d_r}}}
\end{equation}
The diffraction contribution is the difference between the height of the center of the Fresnel region and the height of the scatterer, expressed as:
\begin{equation}
{K_{df}} = {c_{df}}\left( {{h_i} - {H_i}} \right)
\end{equation}
where ${c_{df}}$ is the knowledge coefficients of diffraction.

Furthermore, unlike previous works, which distinguish the presence of obstruction, the degree of blockage caused by the scatterers during propagation is quantified.
The blockage contribution is represented as the ratio of the distance between the blockage scatterer center point to the line connecting TX and RX to the line connecting the bottom of the blockage scatterer, multiplied by a knowledge coefficient, expressed as:
\begin{equation}
{K_{block}} = \frac{{{{\bf{P}}_i} \cdot \overrightarrow {{{\bf{P}}_t}{{\bf{P}}_r}} }}{{||\overrightarrow {{{\bf{P}}_t}{{\bf{P}}_r}} ||}} \cdot \frac{{{c_{block}}}}{{\sqrt {{l_i}^2 + {w_i}^2} }}
\end{equation}
where $l$ and $w$ represent the length and width of the scatterer, respectively.

\subsection{Knowledge coefficient}\label{3.5}

In the case of multiple scatters from the BS to the user terminal UT in the communication link, there is a difference in the relative distance between the scatters and the UT.
Although location information quantifies the propagation contribution, from the perspective of an environment sensing device, the distance at which the same scatterer reaches two very close RX is almost the same.
However, the signal loss caused by the scatterer is significantly different from the perspective of the received signal at RX. 
For example, assuming that we have obtained effective scatterers, only a few of these scatterers may provide significant signal gain. 
Therefore, these scatterers that provide the main gain need to be given appropriate weights based on prior knowledge.

The knowledge coefficients ${\bf{C}} = \{ {c_{diag}},{c_{ref - i}},{c_{ref - g}},{c_{df}},{c_{block}}\} $ denotes the upper bound for impending blockage.
When dis > c1, the communication link from BS to the UT is in a complete openness scenario.
${c_{ref - i}}$ to ${c_{block}}$ are the knowledge coefficients for scatter reflection, ground reflection, diffraction, and blockage, respectively.

By combining prior knowledge with grid search, we set specific knowledge coefficients. 
For the upper bound of impending blockage, a grid search method is employed with a step size of 0.05 within the range of $\left( {1/2,3/2} \right]$, resulting in the selection of the knowledge coefficient as ${c_{diag}}$ = 0.75.
For the reflection knowledge coefficients, analysis of channel simulation data based on ray tracing provides the following prior knowledge: 
(1) For reflection points on the same scatterer, the closer the reflection point is to the center point of the scatterer, the greater the powers in the reflection path.
(2) For reflection points on different scatterers, the shorter the sum of distances from the reflection point to the BS and UT, the greater the powers in the reflection path.
(3) In scenarios of impending blockage and complete blockage, the powers of ground reflection paths are minimal, prioritizing the quantification of scatterer reflection paths.
(4) In scenarios of complete openness, the powers of ground reflection paths are significant. However, in the presence of scatterers in the vicinity, scatterers still induce greater losses.

Based on these findings, the scatterer numbers are sorted in descending order based on the sum of distances from the reflection point to the BS and UT.
Assign an initial value of ${c_{ref - i}}$ = 5 and decrement weights by 0.2, set ${c_{ref - g}}$ = 0.5. 
Since diffraction and blockage contribution are quantified on the only blockage scatterer, set ${c_{df}}$ = ${c_{block}}$ = 1.

Based on stochastic geometry and electromagnetic propagation theory to acquire propagation knowledge, integrating prior experience and grid search methods to set knowledge coefficients aims to enhance interpretability between the environment and reflections, diffraction, and blockages.

\section{Simple structure CNN-based path loss prediction method}

The constructed REK enhances interpretability, enabling satisfactory channel parameter prediction through a simple neural network mapping. 
To validate the constructed REK, we employ a simple AI predictor for the path loss prediction task.
The simple structure weakens the neural network's feature extraction ability and makes the prediction result more dependent on the knowledge provided by the REK.

\begin{figure}[tbh]
	\centering
	\includegraphics[scale=0.3]{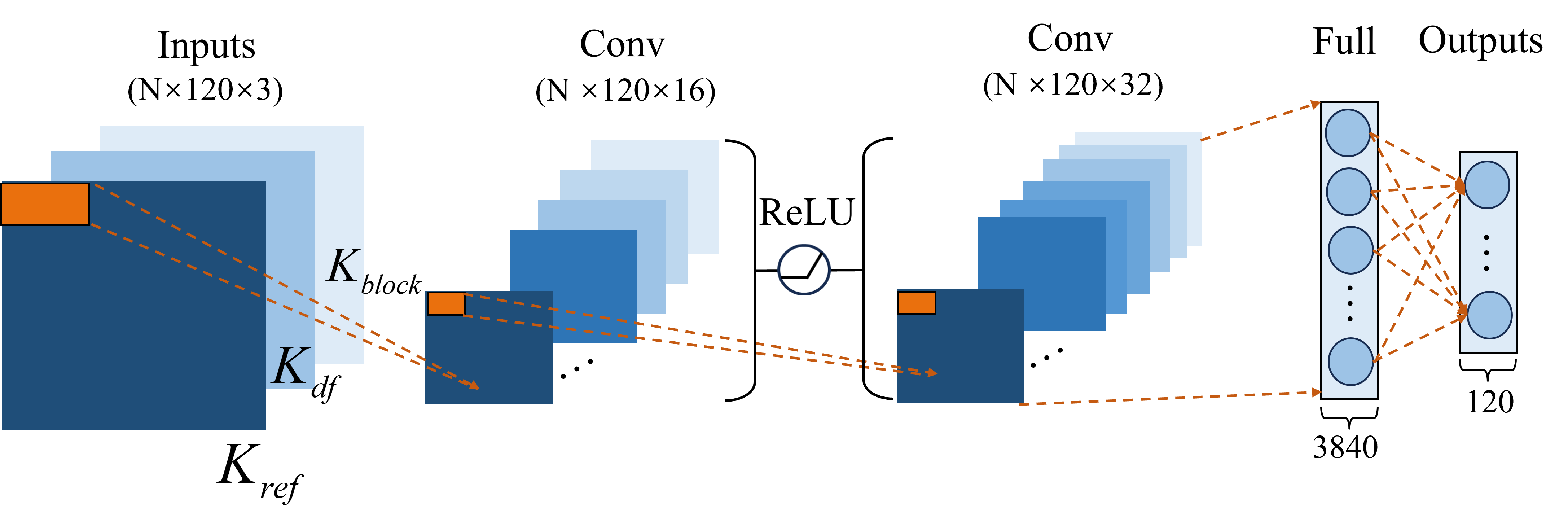}
	\caption{The lightweight CNN with only two convolutional layers for path loss prediction task.} \label{CNN}
\end{figure}

CNN is a neural network that handles mesh-like topological data structures \cite{li2021survey}. 
The considered input data shape consists of $I$ columns of continuous trajectories, with each column containing $J$ location points and each location point containing the REK matrix, making it suitable as input data for CNNs.
This network consists of convolutional, pooling, and fully connected layers. 
The convolutional and pooling layers form multiple convolutional blocks that extract features from the input. The proposed network structure includes two convolutional layers and a fully connected layer, as shown in Fig \ref{CNN}.

The primary function of the convolutional layer is to perceive local features of the input and transform these local features into global representations, expressed as:

\begin{footnotesize}
\begin{equation}
S(i, j) = (X * W)(i, j) = \sum_{m}\sum_{n} X(m, n)W(i-m, j-n)
\end{equation}
\end{footnotesize}
where $X$ is the REK of multiple RX locations, $W$ is the convolutional kernel, $m$ and $n$ are the indices for the convolution operation.
The input and output data channels are 3 and 16,  with a 3 $\times$ 3 convolution kernel. The second convolutional layer has 512 neurons, with input and output data channels of 16 and 32, respectively.
Nonlinear transformations of data were conducted using the ReLU activation function between the two convolutional layers.
The fully connected layer unfolds the feature maps extracted by convolutional layers into a one-dimensional vector. This vector is then transformed through a matrix multiplication with the weights $W$ and an addition of the bias $b$, producing the final output, expressed as $y = f(Wx + b)$.
The input to the fully connected layer is a 120 $\times$ 3 feature matrix.

\section{REK verification}
A quantitative evaluation for REK and a performance comparison based on the proposed lightweight CNN predictors are conducted.
This section reports some representative numerical results in terms of a knowledge graph of propagation contribution, scatterers selection and prediction accuracy, and training time. 

\subsection{Datasets and simulation configuration}

All datasets are generated by the Beijing University of Posts and Telecommunications and China Mobile Communications Group-DataAI - 6G Dataset (BUPTCMCC - DataAI - 6G Dataset) \cite{shen2023dataai}, which can download them from 
https://hpc.bupt.edu.cn/dataset-public/datasets/9. 
From the 646 m $\times$ 290 m simulation scene to generate channel data from the TX to RX deployment area, obtain path loss, and simulate path files. 
The size of the RX2 deployment area is 59.5 m $\times$ 30 m. 
Starting from the bottom left corner of the RX deployment area, every 61 RXs form a row with a spacing of 0.5m, totaling 120 rows arranged to the top right corner, totaling 7320 RXs. The specific simulation configurations are shown in Table \ref{dataconfig}.

\begin{table}[thp]\footnotesize
\centering
\caption{Simulation configuration} \label{dataconfig}
\renewcommand\arraystretch{1} 
\addtolength{\tabcolsep}{10pt} 
\begin{tabular*}{7.5cm}{cc} 
	\toprule[0.75pt]
	 \textbf{Parameter}     & \textbf{Values}  \\
	\midrule[0.5pt]
          Scenario size     &   646 m $\times$ 290 m  \\
	   Frequency     &  3.5 GHz  \\
	   Bandwidth     &  200 MHz  \\
          TX antenna type     &  Omnidirectional  \\
	   TX antenna height   &  20 m  \\
	   RX deployment size   &  59.5 m $\times$ 30 m  \\
	   RX antenna type     &  Omnidirectional  \\
	   RX antenna height     &  1.5 m \\
          RX deployment spacing     &  0.5 m  \\
	\bottomrule[0.75pt]
\end{tabular*}
\end{table}

\begin{figure}[tbh]
	\centering
	\includegraphics[scale=0.65]{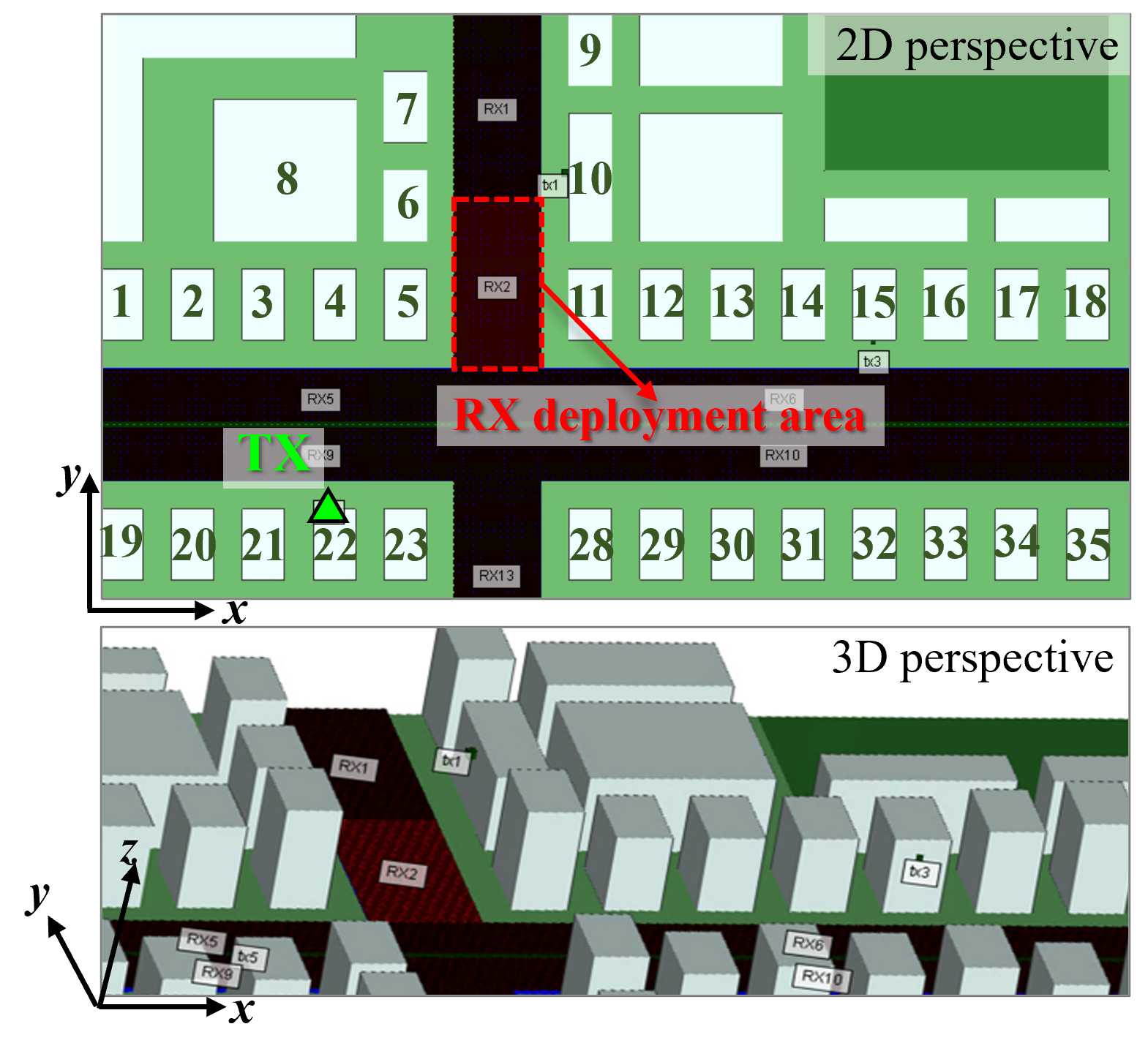}
	\caption{2D and 3D perspective of the simulation scene, scatterer numbers, TX location and RX deployment area.} \label{scene2Dand3D}
\end{figure}

Environment information is obtained from the simulation file, including the three-dimensional (3D) coordinates of TX and RX and each scatterer's maximum and minimum 3D coordinates. 
The center point coordinates of the scatterers are calculated based on the maximum and minimum 3D coordinates. 
The simulation scenario's two-dimensional (2D) and 3D perspectives are shown in Fig. \ref{scene2Dand3D}, providing labels for each scatterer to evaluate the accurate selection of effective scatterers.
There are a total of 7320 data samples, each with one propagation knowledge matrix that includes three types of propagation: reflection, diffraction, and blockage.
75\% of the data is allocated for training, and the remaining 25\% is testing data.
Table \ref{CNNconfig} shows the specific network hyperparameters configuration configuration.
All simulation results are uniformly tested on devices with i5-14600kf (32G) configurations and Nvidia RTX 4080 (16G) and calculated more than ten times to take the average.

\begin{table}[thp]\footnotesize
\centering
\caption{Hyperparameters configuration} \label{CNNconfig}
\renewcommand\arraystretch{1} 
\addtolength{\tabcolsep}{15pt} 
\begin{tabular*}{7cm}{cc} 
	\toprule[0.75pt]
	 \textbf{Parameter}     & \textbf{Values}  \\
	\midrule[0.5pt]
          Batch size     &   16 \\
	   Cost function     &  NRMSE  \\
	   Actuation function     &  ReLU  \\
          Learning rate     &  0.001  \\
	   Train data number   &  5760  \\
	   Test data number   &  1560  \\
          Hiden neurons   &  16, 32, 120  \\
	\bottomrule[0.75pt]
\end{tabular*}
\end{table}

A training process starts from an initial state where all weights and biases are randomly selected. 
Normalized root mean squared  error (NRMSE), a metric for measuring prediction accuracy, is also chosen as the cost function here for training, given by
\begin{equation}
NRMSE = \sqrt {\frac{{\sum\limits_{l = 1}^{{L_{tes}}_t} {{{\left( {{Y_{real}}(t) - {Y_{pred}}(t)} \right)}^2}} }}{{{L_{test}}{\sigma ^2}}}} 
\end{equation}
where ${{L_{test}}}$ is the sample number of the constructed REK matrices, ${{Y_{real}}(t)}$ and ${{Y_{pred}}(t)}$ are the real and predicted output values at the other location $l$, respectively, ${\sigma ^2}$ is the variance of ${{Y_{pred}}(t)}$.
The training process iterates until the network completes the specified training epochs or meets the expected convergence condition. 
Once completed, the trained network can predict path loss at other locations.

\subsection{REK generation and evaluation}

\begin{figure*}[tbh]
	\centering
	\includegraphics[scale=0.45]{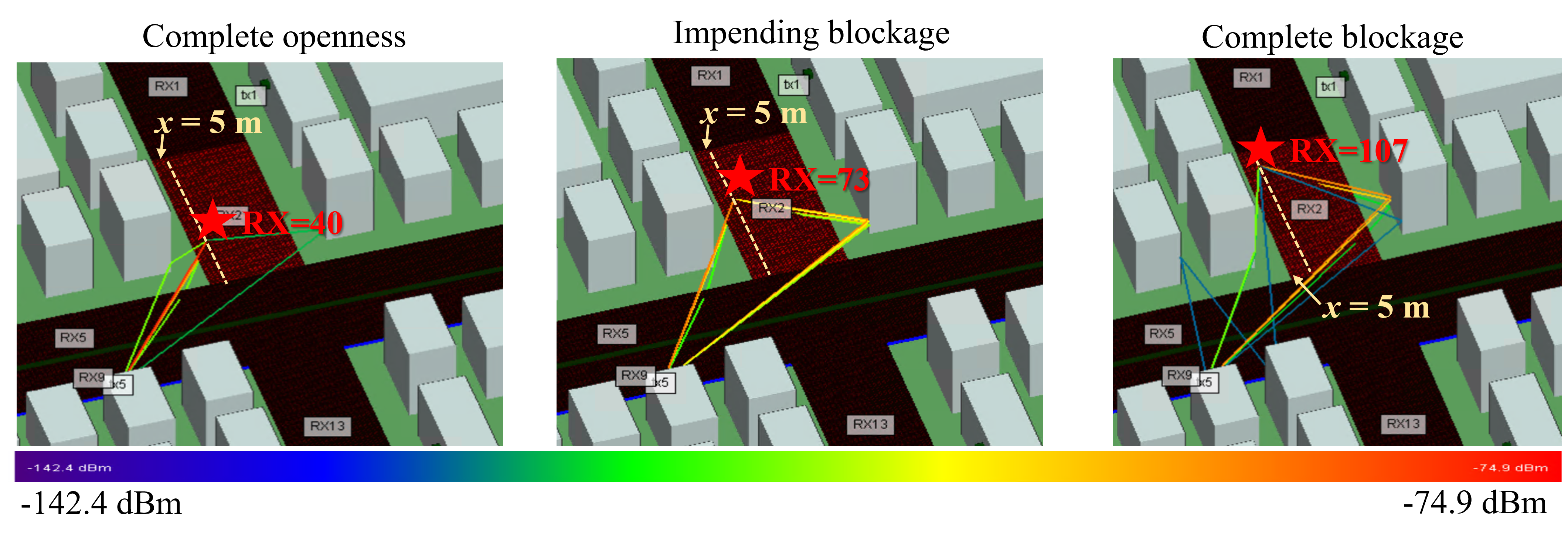}
	\caption{Scatterers and main path links passing from TX to RX under the condition of
 complete openness, impending blockage, and complete blockage (number of paths =10, $x$ = 5 m, $x$ is the horizontal distance from the left roadside).} \label{RTscene}
\end{figure*}

First, verifying whether the constructed REK can correspond to the effects of reflection, diffraction and occlusion in the actual scene is necessary.
The simulated paths obtained by RT are approximated to the actual paths in this paper. 
Fig. \ref{RTscene} shows the scatterers traversed by the first ten paths arranged in reverse order of power under three scenarios: complete openness, impending blockage, and complete blockage.
The color of the paths indicates the received power at the RX, with closer proximity to red indicating higher received power. 
The RX labels associated with the three scenarios are RX=40, 73, and 107.

Table \ref{SelectAccuray} presents the accuracy of effective scatterer selection based on the ellipsoid model using stochastic geometry. 
Considering that scatterers closer to the RX have a greater impact, we propose a scatterer selection accuracy evaluation metric with uniformly decreasing weights, expressed as:
\begin{equation}
\begin{array}{l}
A = \sum\limits_{i \in {S_{select}}} {{w_i}} ,\\
{w_i} = \left\{ \begin{array}{l}
0.3 - \left( {n - 1} \right) \times 0.5,i = S_{{\rm{real}}}^{(n)}\\
0,i \in {S_{real}},i \ne S_{{\rm{real}}}^{(n)}\\
 - 0.1,i \notin {S_{real}}
\end{array} \right.
\end{array}
\end{equation}
where ${S_{\text{select}}}$ is the effective scatterers labels selected by the proposed method, and ${S_{\text{real}}}$ is the scatterer labels traversed by the first 25 paths obtained by the RT method, sorted by receiving power in reverse order, $S_{{\rm{real}}}^{(n)}$ is the first five values in the ${S_{\text{real}}}$, $n=5$.
When the proposed method assigns scatterer indices corresponding to one of the first five values in the ${S_{\text{real}}}$, a weighting scheme is applied starting with a weight of 0.3 for the first value, decreasing sequentially by 0.05 for the second to fifth values.
Weights of 0 are assigned to scatterers that are in the ${S_{\text{real}}}$ but not among the $S_{{\rm{real}}}^{(5)}$, while scatterers not belonging to ${S_{\text{real}}}$ are assigned a weight of -0.1. The final selection accuracy is obtained by summing these weights.
In the cases of impending blockage and complete blockage, the accuracy of identified scatterers is at 90\%, with a slight decrease in accuracy for complete openness. 
In LoS conditions, the majority of the path received power comes from the direct path, where accurately selecting the majority of scatterers providing sub-path power does not significantly impact subsequent knowledge construction.

\begin{table*}[!t]
\centering
\addtolength{\tabcolsep}{4.2pt} 
\footnotesize 
\renewcommand\arraystretch{1.25}  
\caption{Accuracy of obtaining scatterers based on geometric construction method.}\label{SelectAccuray}
\begin{tabular}{ c c c c c}  
	\toprule[0.75pt]
\multirow{2}*{RX number} & \multicolumn{2}{c}{Scatterers number} & \multicolumn{1}{c}{\textbf{Accuracy}} \\ \cline{2-3}
                     & {Proposed method} & {RT method(Power reverse order)} & \textbf{(\%)}\\  
\midrule[0.5pt]
       {40} & {4, 5, 23} & {5, 11, 23, 4, 10, 6, 2, 22}  & \textbf{65}\\
\midrule[0.5pt]
       {73} & {4, 5, 11, 23} & {5, 11, 23, 4, 10, 6}  & \textbf{90}\\
\midrule[0.5pt]
       {107} & {4, 5, 6, 8, 11, 23} & {11, 6, 5, 4, 23, 10, 27, 28, 29, 7}  & \textbf{90}\\       
\bottomrule[0.75pt]
\end{tabular}
\end{table*}

\begin{figure}[tbh]
	\centering
	\includegraphics[scale=0.31]{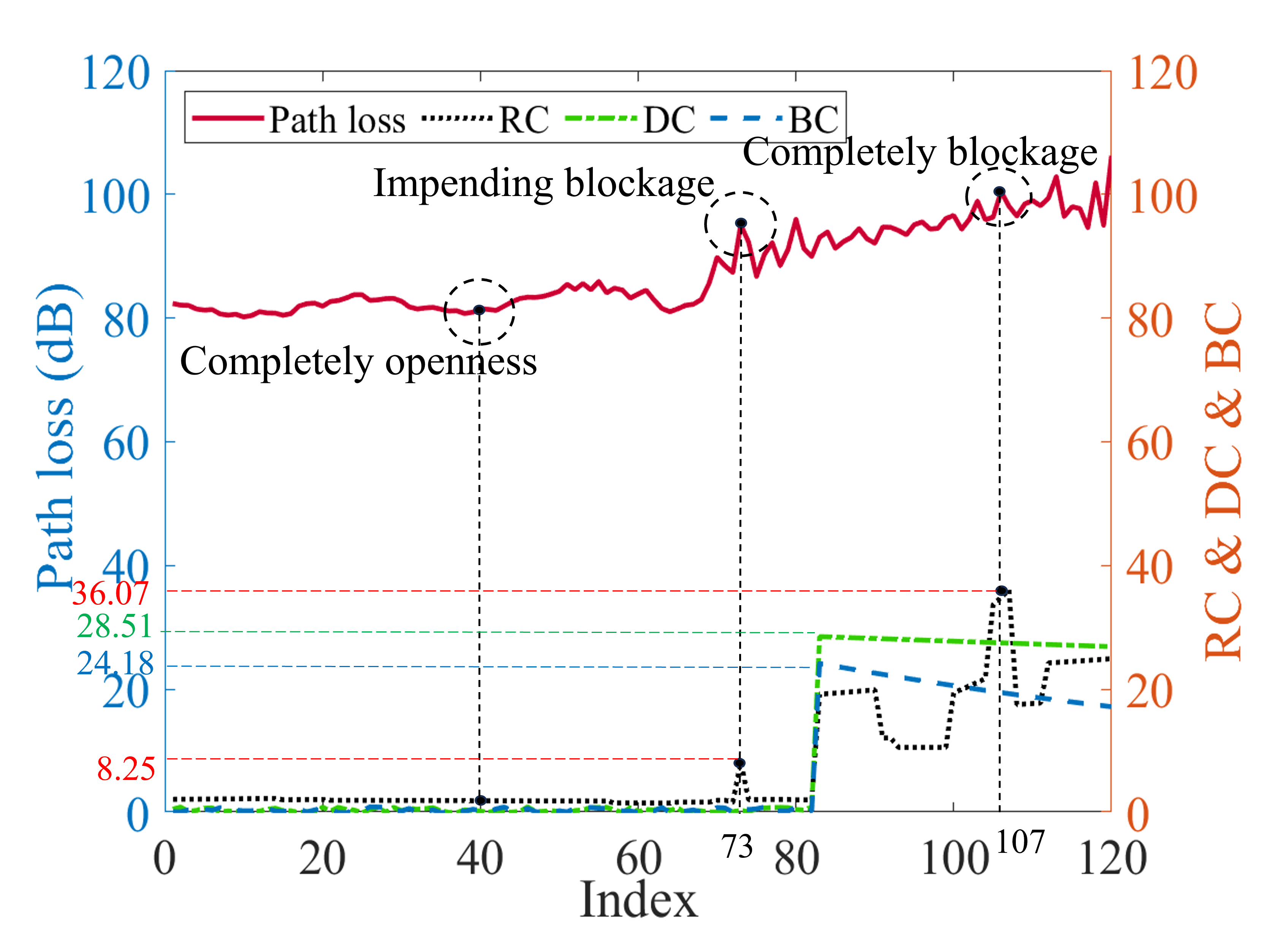}
	\caption{Path loss curve and the reflections, diffraction, and blockage contributions under the corresponding location ($x$ = 5 m).} \label{plot1}
\end{figure}

\begin{figure}[tbh]
	\centering
	\includegraphics[scale=0.35]{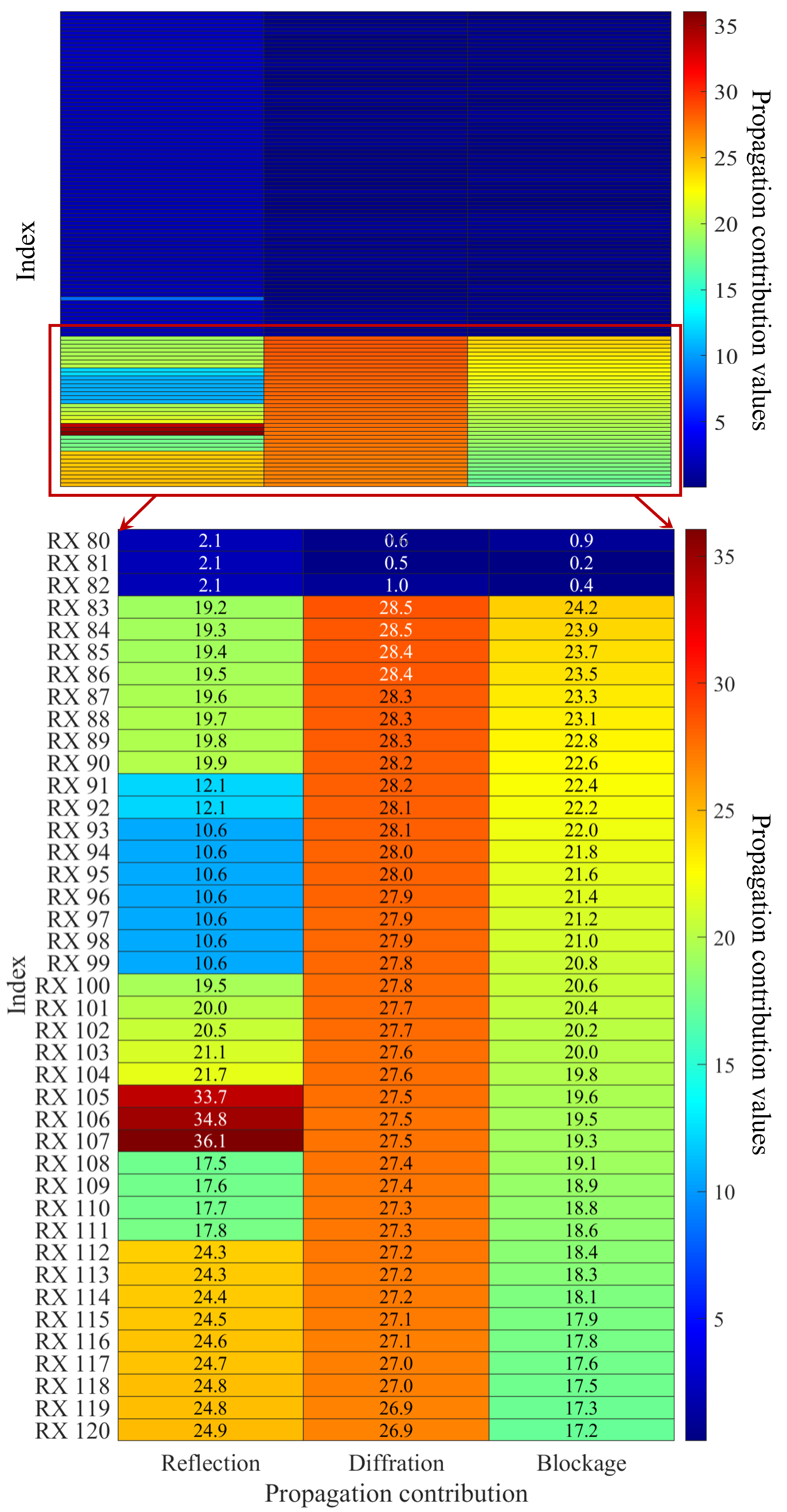}
	\caption{Radio environment knowledge spectrum for reflection, diffraction and blockage propagation ($x$ = 5 m).} \label{heatmap}
\end{figure}

Fig. \ref{plot1} shows the dual y-axis plot of REK corresponding to the path loss curve locations in Fig. \ref{RTscene} scenario, where the left y-axis is path loss and the right y-axis denotes contributions from reflection, diffraction, and blockage. 
The x-axis indicates the location labels of the RX. 
From the whole perspective, when RX < 80, the impact of REK is minimal, whereas for RX > 80, contributions from reflection (RC), diffraction (DC), and blockage (BC) become more intricate.
In the region where RX ranges from 0 to 60, representing a completely open scenario, the path loss remains stable with minimal propagation contributions. Most received power originates from the direct path, resulting in low path loss. 
At RX = 73, a notable increase in reflection contribution is observed with ${K_{ref}}$ = 8.25, capturing the reflective effects introduced by surrounding scatterers. 
In complete blockage scenarios with RX > 80, the maximum reflection contribution peaks at 36.07, occurring at RX = 107. 
The maximum values for diffraction and obstruction contributions are 28.51 and 24.18, respectively, and they exhibit a gradual decrease trend.
However, a significant increase is observed compared to complete openness and impending blockage scenarios.
Furthermore, in conjunction with Fig. \ref{RTscene}, it is evident that at RX = 107, the reflection paths are complex, with the total power from reflection paths significantly surpassing that from diffraction paths.
Thus, it confirms that the proposed REK construction method effectively characterizes the contributions of different propagation types to path loss at various locations.
The complete REK quantification spectrum along this trajectory is illustrated in Fig. \ref{heatmap}, with an enlarged subgraph depicting the portion where RX > 80. In subsequent predictions, the REK spectrum for each trajectory is utilized as input.

\subsection{Prediction performance }

\begin{figure}[!htb]\small
\centering
\begin{tabular}{c}
\includegraphics[scale=0.25]{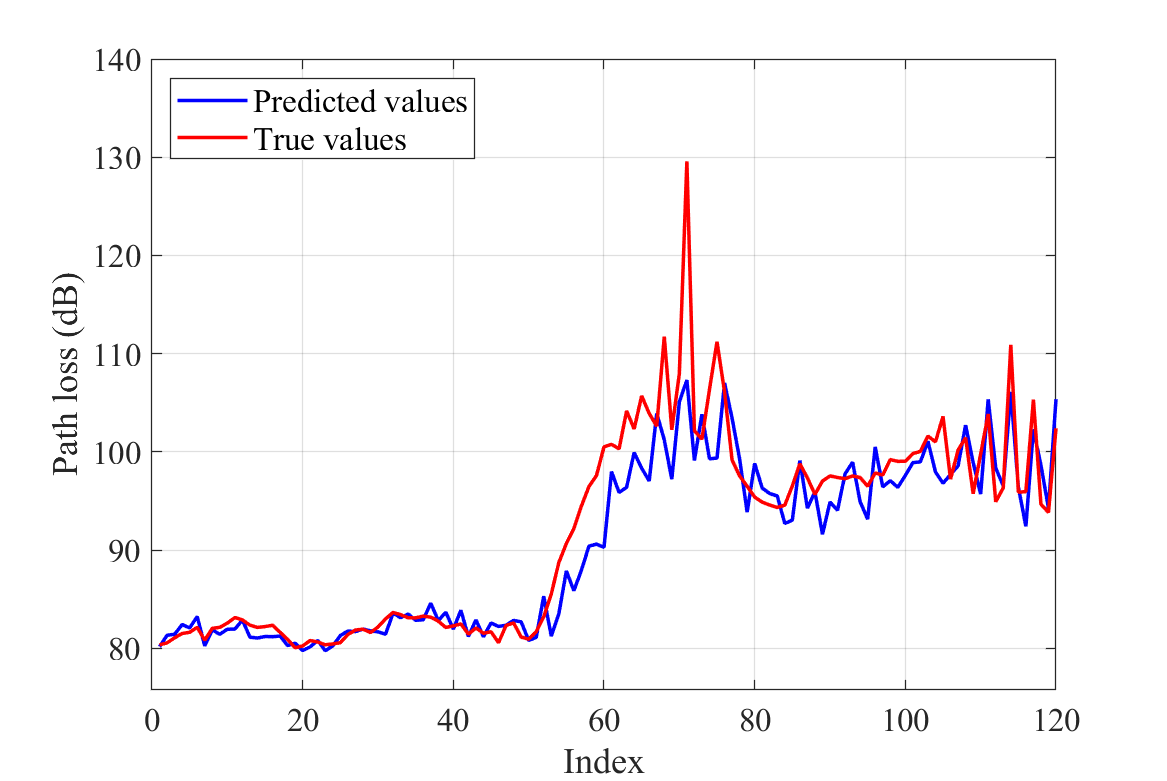}\\
{\footnotesize\sf (a)} \\[3mm]
\includegraphics[scale=0.25]{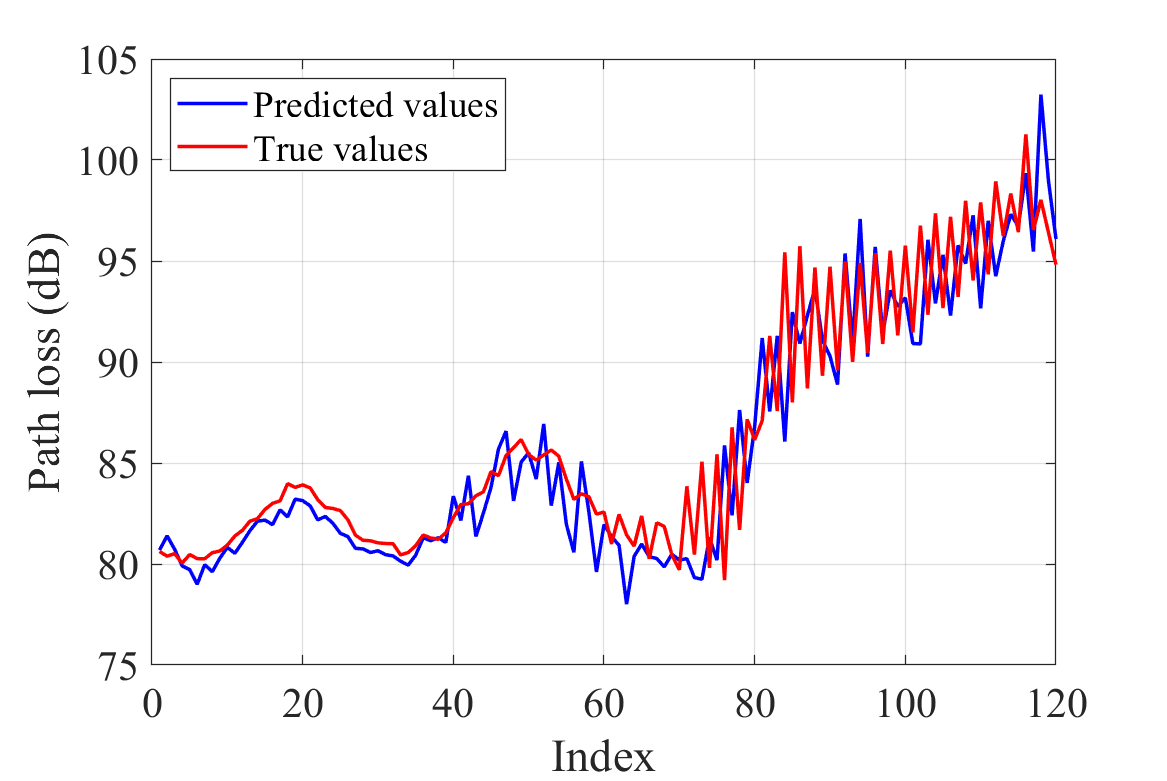}\\
{\footnotesize\sf (b)} \\
\end{tabular}
\caption{The comparison curve between the true and predicted values based on AI and the REK construction. (a) $x$ = 25 m, (b) $x$ = 30 m}
\label{Preandreal}
\end{figure}

The predicted values generated by the proposed knowledge and AI-based predictor are compared and evaluated against the path loss derived from RT as an approximation of true values.
The training data is as little as possible, achieving an NRMSE = 0.3 and training and test time of 5 and 0.004 seconds, respectively.
Fig. \ref{Preandreal} gives the predicted and true values comparison curves, obtained by knowledge and CNN-based method for the horizontal distance at 25 m  and 30 m from the left roadside. 
When x = 25 m, the trajectory is located in the middle of the road, and the buildings on both sides affect the radio waves. 
When x = 30 m, the trajectory moves towards the right side of the road, potentially experiencing more complex propagation types influenced by the buildings on the right roadside.
As this figure shows, the trained model precisely captures the variations in path loss at different locations, particularly in challenging scenarios such as complete blockage. 
Combining with Fig. \ref{RTscene}, it is evident that when x = 30 m, nearing the right roadside, the radio experiences a diverse range of propagation types influenced by complex scatterers. 
The proposed approach effectively aligns with the true path loss trends in such instances. 
Under complete openness scenarios, as shown in Fig. \ref{Preandreal} (a) where index = 0 - 60, the predicted fitting curve almost coincides with the true values.

\begin{figure}[!htb]\small
\centering
\begin{tabular}{c}
\includegraphics[scale=0.25]{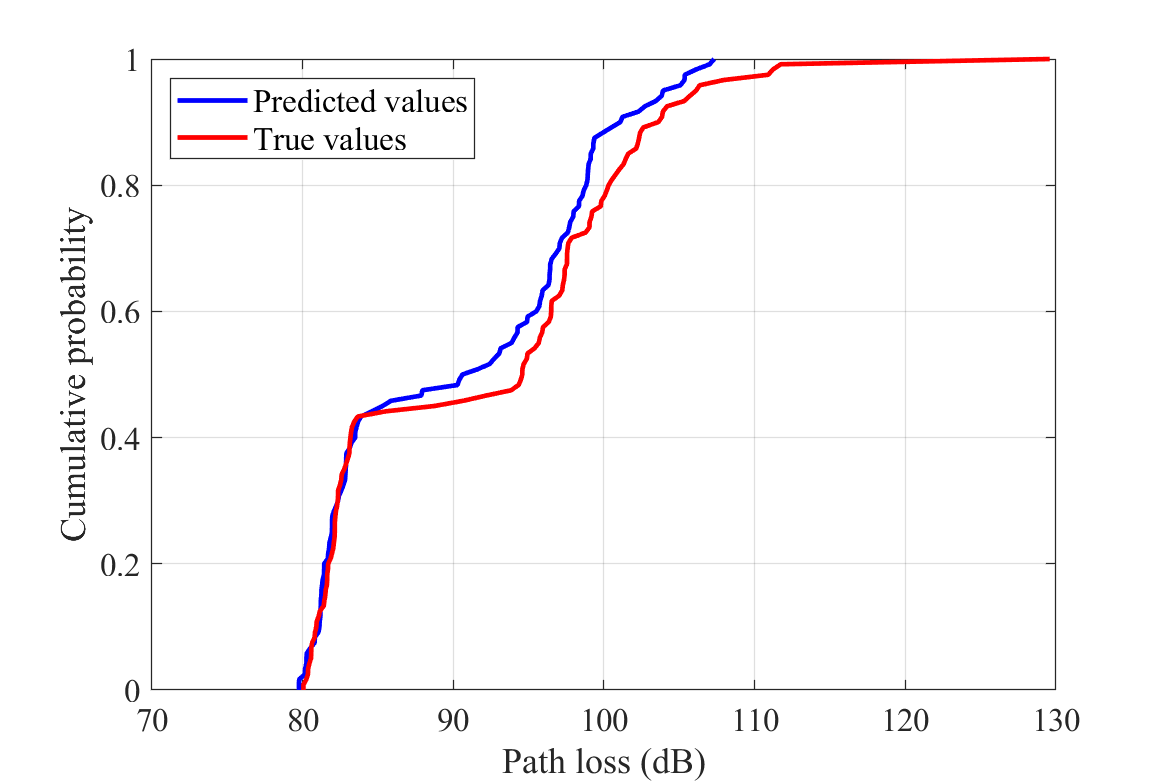}\\
{\footnotesize\sf (a)} \\[3mm]
\includegraphics[scale=0.25]{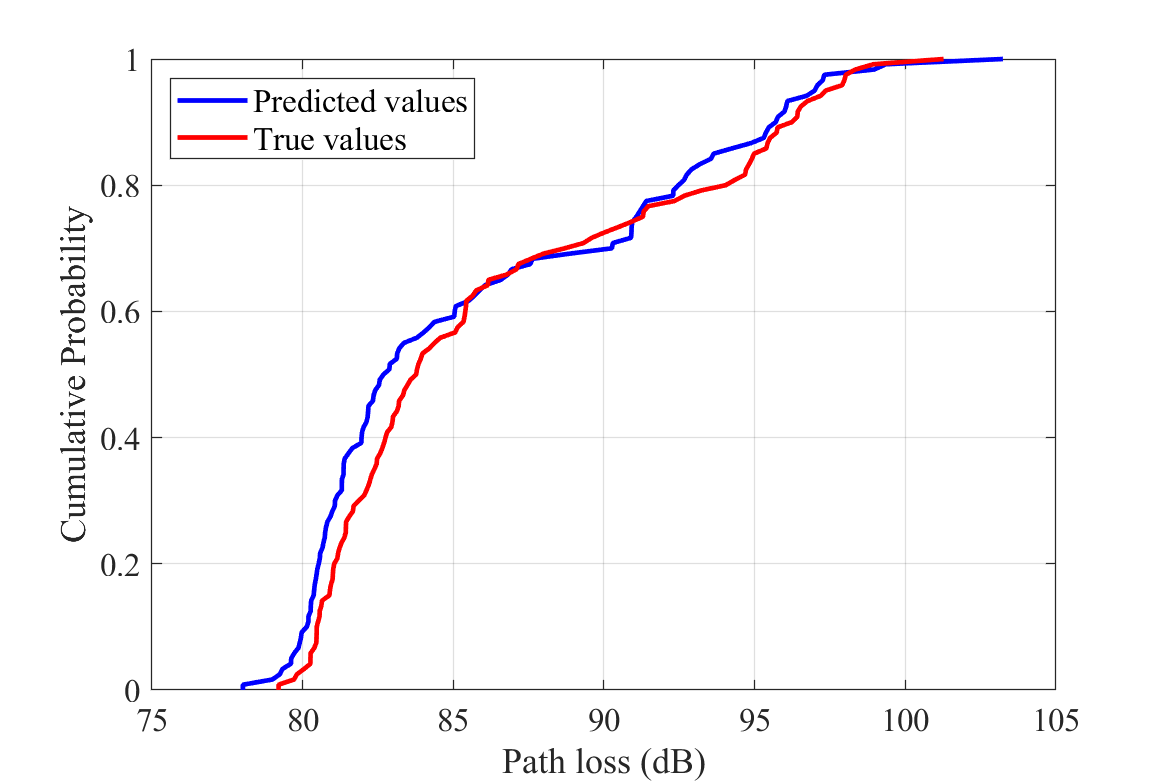}\\
{\footnotesize\sf (b)} \\
\end{tabular}
\caption{CDF plots of the true and the predicted outputs of knowledge and CNN-based in path loss prediction tasks. (a) $x$ = 25 m, (b) $x$ = 30 m}
\label{cdfPic}
\end{figure}

\begin{table*}[!t]
\centering
\addtolength{\tabcolsep}{4.2pt} 
\footnotesize 
\renewcommand\arraystretch{1.2}  
\caption{Quantification of box plot for knowledge and CNN-based method in path loss prediction tasks.}\label{table4}
\begin{tabular}{ c c c c c c c c}  
	\toprule[0.75pt]
 x & Output & UQ & LQ & UB & LB & MED & OL \\ \cline{1-8}
 
\multirow{2}*{25 m} & Predicted & 98.9719 & 93.1079 & 81.9457 & 123.0111 & 57.9065 & 0 \\ \cline{2-8}

                     & True & 99.1884 & 94.5806 & 82.1302 & 124.7756 & 56.5430 & 1 \\ 
\midrule[0.5pt]
\multirow{2}*{30 m} & Predicted & 91.1302 & 82.7795 & 80.7615 & 106.6832 & 65.2084 & 0 \\\cline{2-8}
                     & True & 91.3047 & 83.7807 & 81.4328 & 106.1127 & 66.6249 & 0 \\      
\bottomrule[0.75pt]

\end{tabular}
\end{table*}

From the view of statistical analysis, the nonlinear approximation of the knowledge and AI-based prediction model is further evaluated, including cumulative distribution function (CDF) and quantitative statistical values comparison.
In Fig. \ref{cdfPic}, the effectiveness of the proposed knowledge and AI-based model is validated by comparing the CDF plots of the predicted and true values.
The proposed model captures the complete process of path loss resulting from the superlocation of different propagation types when nearing the right roadside (x=30), as shown in Fig. \ref{cdfPic}(b). 
The predicted values exhibit similar numerical values and proportions to the true values, falling within the same distribution range, albeit not perfectly overlapping. 
Furthermore, the model did not predict the portion where PL > 110 dB, as shown in Fig. \ref{cdfPic} (a). This segment displays sudden spikes in one to two path loss values, collectively identified as outliers in path loss.
This phenomenon corresponds to Fig. \ref{Preandreal} (a).
Table \ref{table4} presents the quantitative statistical data of predicted and true path loss, including the upper and lower quartiles (UQ and LQ), the upper and lower bound (UB and LB), median values(MED) and outliers (OL).
The specific data points differ by less than 1, which verifies the validity of the constructed REK.

\section{Conclusion}

Constructing the relationship between the environment and the channel is crucial for mapping real-world physical channels to generate DTC in the digital world. 
In this paper, we propose a method of REK construction inspired by electromagnetic wave properties to solve the high complexity of the construction process caused by the redundancy of environment information and the ambiguity of relationship correlation.
Firstly, an effective environment information range selection scheme based on random geometry is proposed to reduce redundancy in environmental information. 
Subsequently, the contributions of electromagnetic wave reflection, diffraction, and the blockage of scatterers in the NLoS scenario are quantified using environment information. 
Based on the concept of REKP, which utilizes environment information to construct essential relationships and form REK, a process for implementing REK construction is proposed to replace part of the complex feature extraction process of neural networks. 
Experiments on a path loss prediction task based on a lightweight CNN are conducted to validate REK's effectiveness. 
The results show that using REK as input maintains a prediction error of 0.3 with only two convolutional layers in the network structure, with a testing time of only 0.04 seconds, effectively reducing network complexity.

\end{document}